# Detection of poloidal magnetic flux emission from a plasma focus.

S K H Auluck

International Scientific Committee for Dense Magnetized Plasmas,

http://www.icdmp.pl/isc-dmp

Hery 23, P.O. Box 49, 00-908 Warsaw, Poland

email: skhauluck@gmail.com

Abstract:

Direct experimental evidence for the existence of axial magnetic field both in the radial implosion phase and the pinch phase of a plasma focus has raised many questions of fundamental importance. The most fundamental of these is the fact of its existence, which is incomprehensible in terms of a conventional view of plasma physics. The plasma is known to have an axis of symmetry in the radial implosion phase. The axial magnetic field has a defined polarity with respect to the axis. Since the equations of magnetohydrodynamics are unchanged by flipping the sign of the axial coordinate, the polarity of the axial magnetic field must arise from the initial conditions. Exactly how the initial conditions determine the polarity of the axial magnetic field is not clear. More data are needed to guide theoretical developments of this question. There are several technical problems with the conventional techniques of axial magnetic field measurement such as magnetic probes and Faraday rotation measurement. The combined effect of theoretical and experimental difficulties is that researchers avoid research related to the axial magnetic field in axisymmetric plasmas in favour of more result-oriented projects, with the unfortunate consequence that questions related to the presence of axial magnetic field in axially symmetric plasmas remain unresolved. In order to promote research related to the questions related to the axial magnetic field in the plasma focus and other plasmas with an axial symmetry, this paper introduces a new technique that skirts the technical difficulties of measurement by redefining the objectives and scope of measurement. The paper is written in a tutorial format for the benefit of young researchers just beginning their research career and hence will not be published in regular journals. Peer scientists and young researchers are encouraged to contact the author directly.

I. **Introduction:**

Direct experimental evidence for existence of axial magnetic field both in the radial implosion phase and the pinch phase of a plasma focus [1,2,3,4] has raised many questions of fundamental importance [5]. The most fundamental of these is the fact of its existence, which is incomprehensible [6] in terms of a conventional view of plasma physics, as discussed below.

Equations governing the evolution of plasmas come in several sets of differing complexity (ideal MHD, resistive MHD, Hall MHD, two-fluid or three-fluid MHD, radiation MHD, Vlasov fluid, Boltzmann kinetic equation etc.). All of these equation systems are unchanged by flipping the sign of the axial coordinate, because they do not have an inbuilt preferred direction. Since the axial magnetic field has a defined polarity with respect to the axial coordinate, it can only arise from the initial conditions. There are only 4 external vector fields [6] that can provide such a reference direction by way of initial conditions. They are: Earth's magnetic field, Earth's angular momentum, direction of axial current flow and direction of axial plasma velocity. The problem is: it is very difficult to conceive of a mechanism that can generate an axial magnetic field in an axially symmetric plasma starting from any of these initial conditions. Usual dynamo and flux amplification processes fail to reproduce key features of experimental data [6]. More complex processes have been proposed [4,5,6] but they remain just proposals in the absence of substantial supporting experimental evidence.

Such theoretical difficulties would normally have provided a strong motivation for designing experiments to address them. However technical difficulties thwart such efforts. Magnetic probes provide a standard technique that is expected to provide a spatially and temporally resolved local measurement of the magnetic field. However, their data are corrupted by three kinds of errors.

   a. The probe technique is invasive. The presence of this physical foreign body makes the plasma flow around it. Since magnetic flux is partially or wholly trapped in the plasma, its magnitude at the position of the probe is changed from what it would have been in the absence of the probe because of this altered magnetized plasma flow.
   b. The plasma often induces an electrostatic potential on the conducting portion of the probe because of formation of a Langmuir sheath around its body. This causes an extra current to flow into the input impedance of the oscilloscope, which may be spuriously ascribed to the rate of change of magnetic flux at its location.
   c. Although the magnetic probe is designed to detect one out of three components of the local magnetic field, in practice it has a non-zero sensitivity to the other two components as well. In the case of the plasma focus, the azimuthal magnetic field is much larger than the axial magnetic field and the cross-contamination from the former can be significant compared to the desired signal from the axial component.

A discussion of efforts to mitigate these sources of error is presented elsewhere [5].

Another standard technique is measurement of the Faraday rotation of the plane of polarization of a laser beam travelling along the magnetic field, in this case along the axis. This was used in the Frascati 1 MJ plasma focus [4] to conclude that an axial magnetic field exists. However, this measurement provides a result that is proportional to the product of electron density and magnetic field integrated along the path of the beam. It therefore does not produce interpretable data of sufficient value to justify the expense and the effort.

Clearly, a new technique is needed to look into the question of existence of poloidal magnetic field that (1) is not too resource-intensive (2) generates new credible information about the physical phenomena involving poloidal magnetic field and (3) can be easily replicated.

For the subsequent discussion, it is necessary to appreciate the difference between axial magnetic field and poloidal magnetic field. The magnetic field is expressed in cylindrical coordinates as

$$\vec{B} = B_r \hat{r} + B_\theta \hat{\theta} + B_z \hat{z} \tag{1}$$

In the presence of azimuthally symmetry, the derivative of any physical field with respect to the azimuthal coordinate $\theta$ is zero. When the magnetic field is expressed in terms of the magnetic vector potential, this translates to

$$B_r = -\frac{\partial A_\theta}{\partial z}; B_z = \frac{1}{r}\frac{\partial}{\partial r}(rA_\theta) \tag{2}$$

$$B_\theta = \frac{\partial A_r}{\partial z} - \frac{\partial A_z}{\partial r} \tag{3}$$

It is clear from equation (2) that the radial and axial components of magnetic field are both related to the azimuthal component of vector potential $A_\theta$ while the azimuthal component of magnetic field is related to radial and axial components of the vector potential which are in fact related to each other via the Lorentz Gauge Condition

$$\vec{\nabla}\cdot\vec{A} = \frac{1}{r}\frac{\partial}{\partial r}(rA_r) + \cancel{\frac{1}{r}\frac{\partial}{\partial \theta}A_\theta} + \frac{\partial}{\partial z}A_z = -\frac{1}{c^2}\frac{\partial \varphi}{\partial t} \approx 0 \tag{4}$$

The mutual relationship between the axial and radial components of the magnetic field is highlighted by treating them as one field – the poloidal magnetic field, and then the azimuthal magnetic field is termed as the toroidal magnetic field:

$$\vec{B}_p = B_r \hat{r} + B_z \hat{z}; \ \vec{B}_t = B_\theta \hat{\theta} \tag{5}$$

These definitions are clearly applicable and relevant only in the presence of azimuthal symmetry where $\partial_\theta \equiv 0$. Any reference to a poloidal magnetic field implies the hidden assumption of azimuthal symmetry.

Let us now consider a circular curve of radius r in the z=0 plane centred at the origin. The element of length of the curve is given by

$$d\vec{\ell} = (rd\theta)\hat{\theta} \tag{6}$$

Keeping in mind the azimuthal symmetry, the line integral of electric field over this curve is

$$V = \oint \vec{E} \cdot d\vec{\ell} = \oint rd\theta E_\theta = 2\pi r E_\theta \tag{7}$$

By the circulation theorem,

$$V = -\frac{d}{dt}\int \vec{B}.d\vec{S} \tag{8}$$

where $d\vec{S} = (2\pi rdr)\hat{z}$ and the integration is over the area enclosed by the circular curve

Now, $E_\theta \to 0$ as $r \to \infty$ or at a continuous conducting boundary such as a metallic vacuum chamber. From (8), this amounts to conservation of the magnetic flux within an infinite boundary or a continuous conducting boundary. It implies that if poloidal magnetic flux is generated within the plasma, it must "return" outside the plasma, contributing a negative value to the integral in (8). This returning poloidal magnetic flux is, in a sense, an emission of poloidal magnetic flux from the plasma since the poloidal magnetic flux inside the plasma is not accessible for measurement.

For a cylindrically symmetric, current-carrying plasma, the magnetic lines of force will spill out from one end, follow a spiral path winding around the axis outside its finite radius and return to the other end. The flux of this composite 3-D magnetic field through a closed curve surrounding it would, however, depend only on the axial component of magnetic field. This flux would vary with the radius of the curve: the more the radius, the more negative flux it encloses and the closer to zero its value would be.

The magnetic field lines can return outside the plasma because they are composed of both axial and radial components. It is this "returning poloidal magnetic flux" outside the plasma that the new technique aims to measure.

This paper is a tutorial introduction to this technique, which was first proposed in the ICDMP International Video Conference on "Constructing a Pro-active Agenda for the Plasma Focus Community" on April 24, 2020 [8].

Section II presents a tutorial on its theory of measurement. Section III illustrates its practical implementation. Section IV briefly overviews issues related to testing, validation and calibration. Section V closes the paper with a summary and discussion.

## II. <u>Theory of measurement:</u>

The measurement of poloidal flux emitted from a solenoid coil carrying time varying current should be a simple matter of surrounding it with a simple loop of wire (called a diamagnetic loop), measuring the signal across it and integrating it with time. This is just an application of equation (8).

The situation becomes slightly more complex if the source that drives a time-varying current into the coil also raises its electric potential which varies with time. In turn, this coil would induce an electric potential on the measuring loop. What would be measured across the diamagnetic loop would be the net result of the induced electric potential and the potential given by equation (8). If there are two identical loops, surrounding the axis in a clockwise (CW) or counter-clockwise (CCW) sense, the contribution from equation (8) to the signal would have opposite signs for the two. So, half their sum would be a signal that arises from electrostatic induction and half their difference would be the contribution from equation (8).

For a real plasma, the situation is even more complex. The presence of strict cylindrical symmetry cannot be guaranteed. In addition, any real plasma always has a charge density. This charge density is a consequence of its finite size as well as its magnetically driven dynamics. Taking the divergence of the Generalized Ohm's Law, one can easily see that the charge density is given by

$$\rho = -\varepsilon_0 \vec{\nabla} \cdot \left( \vec{v} \times \vec{B} \right) \qquad (9)$$

It is clear that any conductor placed in the vicinity of the plasma is going to have an electric potential induced by the charge density. The pair of diamagnetic loops mentioned above would both have this potential on every point. The result is that the cables connected across these diamagnetic loops have a common mode signal in addition to a differential mode signal.

Now, the input amplifier of an oscilloscope has both a differential-mode gain and a common-mode gain [7] and therefore, the signals from the two diamagnetic loops – identical except for the way they are connected to the cables – would be different. This has indeed been observed in the case of a plasma focus [9]. The separation of observed signals from the charge on the plasma and from the magnetic flux variation shows [9] structures which cannot be an artifact from any influence outside the plasma.

However, this measurement using two diamagnetic loops has a potential source of error: the common-mode signal has a linkage with the ground-loop circuit of the facility unless the oscilloscopes are electrically disconnected from the ground. This is possible using a battery-based inverter power supply that is isolated from the laboratory ground. This is not a very good technical solution since the influence of accidental changes in the ground loop circuit cannot be completely eliminated affecting the credibility of data.

The best way to eliminate the influence of the signal component arising from the charge on the plasma is to measure it correctly and account for it in the measurement of the signal from the poloidal magnetic flux emission. The way to do this is explained with reference to Fig 1, which shows the schematic of a coaxial capacitive voltage divider

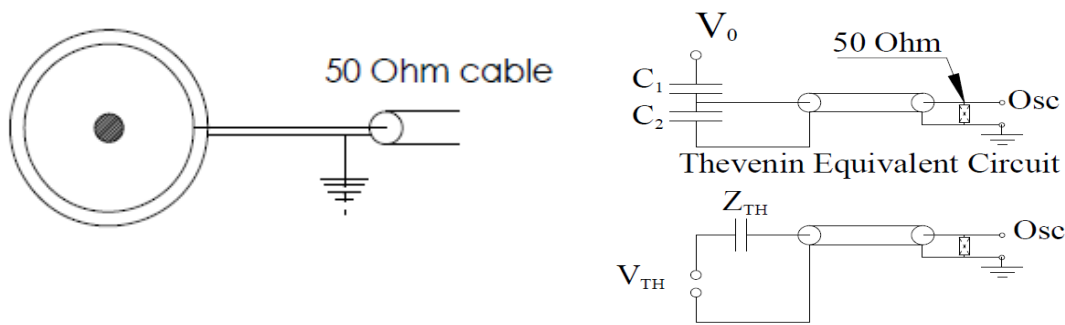

Fig-1. Schematic of a coaxial capacitive voltage probe.

The inner shaded circle represents a high voltage conductor at voltage $V_0$ – it could be a metallic object or a plasma. It is surrounded by two coaxial metallic cylinders isolated from each other. The inner cylinder forms a capacitance $C_1$ with respect to the high voltage point inside and $C_2$ with respect to the outer cylinder. Obviously, $C_1 \ll C_2$. Since the inner cylinder is metallic, the electric field lines emitted from the high voltage point get terminated on it and do not reach the outer conductor. The potential of the outer conductor is then governed by its connection to the ground, which could be the oscilloscope ground. Applying Thevenin's Theorem, the cable would see a voltage source

$$V_{TH} = V_0 \frac{C_1}{C_1 + C_2} \tag{10}$$

in series with an impedance

$$Z_{TH} = Z_{C_1} \| Z_{C_2} = \frac{Z_{C_1} \cdot Z_{C_2}}{Z_{C_1} + Z_{C_2}} = \frac{\frac{1}{j\omega C_1} \cdot \frac{1}{j\omega C_2}}{\frac{1}{j\omega C_1} + \frac{1}{j\omega C_2}} \approx \frac{1}{j\omega C_2} \tag{11}$$

The cable terminated in its characteristic impedance $R_Z$=50 Ohms would then have a current through it given by

$$I_{TH} = \frac{V_{TH}}{Z_{TH} + R_Z} \tag{12}$$

and the voltage across the input of the oscilloscope would be

$$V_{Osc} = R_Z I_{TH} = V_{TH} \frac{R_Z}{Z_{TH} + R_Z} \approx V_0 \frac{C_1}{C_2} \frac{j\omega C_2 R_Z}{1 + j\omega C_2 R_Z} \tag{13}$$

Usually, the condition $j\omega C_2 R_Z \ll 1$ is satisfied. Under this condition, (13) implies

$$V_{Osc} \approx C_1 R_Z \frac{dV_0}{dt} \tag{14}$$

In other words, the signal at the oscilloscope input is produced by the displacement current in the space surrounding the high voltage terminal flowing through the characteristic cable impedance. This qualitative conclusion remains valid even if the height of the cylinder is reduced to a small value, raising concerns about the effectiveness of shielding of the outer cylinder from the influence of the high voltage terminal. The actual signal recorded on the oscilloscope would be half this value on account of voltage division between the cable impedance and the input impedance of the oscilloscope.

This circuit analysis applies to a situation where the capacitance between the high voltage terminal and the inner cylinder is defined by a fixed geometry. In the case of the plasma focus the high voltage terminal is the moving current sheath, that is undergoing a rapid collapse on the axis followed by a reflection. The oscilloscope signal would be proportional to the rate of change of voltage $V_0$ induced on the cylinder by the charge density. The potential $\varphi(\vec{R},t)$ at a field point $\vec{R}$ induced by a charge density distribution $\rho(\vec{R}',t')$ at source points $\vec{r}'$ is given by the textbook formula for the case of non-radiating fields

$$\varphi(\vec{R},t) = \frac{1}{4\pi\varepsilon_0} \int d^3\vec{R}' \frac{\rho(\vec{R}',t')}{|\vec{R}-\vec{R}'|} \tag{15}$$

The retarded time $t' \equiv t - |\vec{R}-\vec{R}'|/c$ can be taken to be equal to t for the conditions of a moderate-sized plasma focus experiment.

The factor $1/|\vec{R}-\vec{R}'|$ can be expressed as

$$\frac{1}{|\vec{R}-\vec{R}'|} = \frac{1}{\sqrt{|\vec{R}|^2+|\vec{R}'|^2-2|\vec{R}||\vec{R}'|\cos(\gamma)}} = \frac{1}{|\vec{R}|\sqrt{1+\eta^2-2\eta\cos(\gamma)}}$$
$$= \frac{1}{|\vec{R}|}\sum_{n=0}^{\infty}\eta^n P_n(\cos\gamma); \Leftarrow \gamma \equiv \angle(\vec{R},\vec{R}'); \eta \equiv |\vec{R}'|/|\vec{R}| \ll 1 \tag{16}$$

In (16), $P_n(x)$ are Legendre Polynomials. In spherical coordinates,

$$\vec{R}\cdot\vec{R}' = |\vec{R}||\vec{R}'|\{\sin\phi\cos\theta\sin\phi'\cos\theta' + \sin\phi\sin\theta\sin\phi'\sin\theta' + \cos\phi\cos\phi'\}$$
$$= |\vec{R}||\vec{R}'|\{\sin\phi\sin\phi'\cos(\theta-\theta')+\cos\phi\cos\phi'\} = |\vec{R}||\vec{R}'|\cos\gamma \tag{17}$$

Therefore,

$$\cos\gamma = \sin\phi\sin\phi'\cos(\theta-\theta')+\cos\phi\cos\phi' \tag{18}$$

In cylindrical polar coordinates, this becomes

$$\cos\gamma = \frac{rr'}{|\vec{R}||\vec{R}'|}\cos(\theta-\theta')+\frac{zz'}{|\vec{R}||\vec{R}'|} \tag{19}$$

The voltage induced on the inner cylinder of height w, inner radius $R_1$ would be an average of $\varphi(\vec{R},t)$ over the axial coordinate as well as the azimuthal coordinate:

$$V_0 = \frac{1}{2\pi} \int_0^{2\pi} d\theta \frac{1}{w} \int_{-w/2}^{w/2} dz \cdot \varphi(\vec{R},t)$$
$$= -\frac{1}{4\pi\varepsilon_0} \int d^3\vec{R}' \rho(\vec{R}',t') \frac{1}{2\pi} \int_0^{2\pi} d\theta \frac{1}{w} \int_{-w/2}^{w/2} dz \cdot \frac{1}{|\vec{R}-\vec{R}'|} \quad (20)$$

In terms of the integral

$$\mathbb{F}(R_1, r', \theta', z') \equiv \frac{1}{2\pi} \int_0^{2\pi} d\theta \frac{1}{w} \int_{-w/2}^{w/2} dz \cdot \frac{1}{|\vec{R}-\vec{R}'|}$$
$$= \frac{1}{2\pi R_1} \frac{1}{w} \int_{-w/2}^{w/2} dz \cdot \sum_{n=0}^{\infty} \left( \frac{\sqrt{r'^2+z'^2}}{\sqrt{r^2+z^2}} \right)^n \quad (21)$$
$$\times \int_0^{2\pi} d\theta P_n \left( \frac{rr'}{\sqrt{r^2+z^2}\sqrt{r'^2+z'^2}} \cos(\theta-\theta') + \frac{zz'}{\sqrt{r^2+z^2}\sqrt{r'^2+z'^2}} \right)$$

(20) can be written as

$$V_0 = -\frac{1}{4\pi} \int dr' \int r' d\theta' \int dz' \rho(r',\theta',z',t') \mathbb{F}(R_1,r',\theta',z') \quad (22)$$

For a real plasma, the field $\rho(r',\theta',z')$ may be a function of the azimuthal coordinate $\theta'$ for a variety of reasons such as non-symmetric initial conditions resulting in its displacement and/or tilt with respect to the axis, instabilities, etc. The integrand of (22) can in principle be expressed in terms of a Fourier series in $\theta'$, consisting of integer harmonics of sine and cosine functions, including a constant term independent of $\theta'$.

What (22) demonstrates is that the voltage induced on the inner cylinder would be independent of the azimuthal variations of the integrand: the integral over the period functions of $\theta'$ would be zero. The only contribution would come from the constant term in the Fourier series.

In physical terms, the real 3-D electric field distribution can be decomposed into a sum of one distribution that is symmetric with respect to the azimuthal coordinate in the device frame of reference and another that contains all the deviations from azimuthal symmetry. What (22) demonstrates is that the coaxial capacitive probe will have zero sensitivity to the second field. Its signal would be proportional to the rate of change of the first field.

Now consider that the two cylinders are cut and joined together as shown in Fig 2.

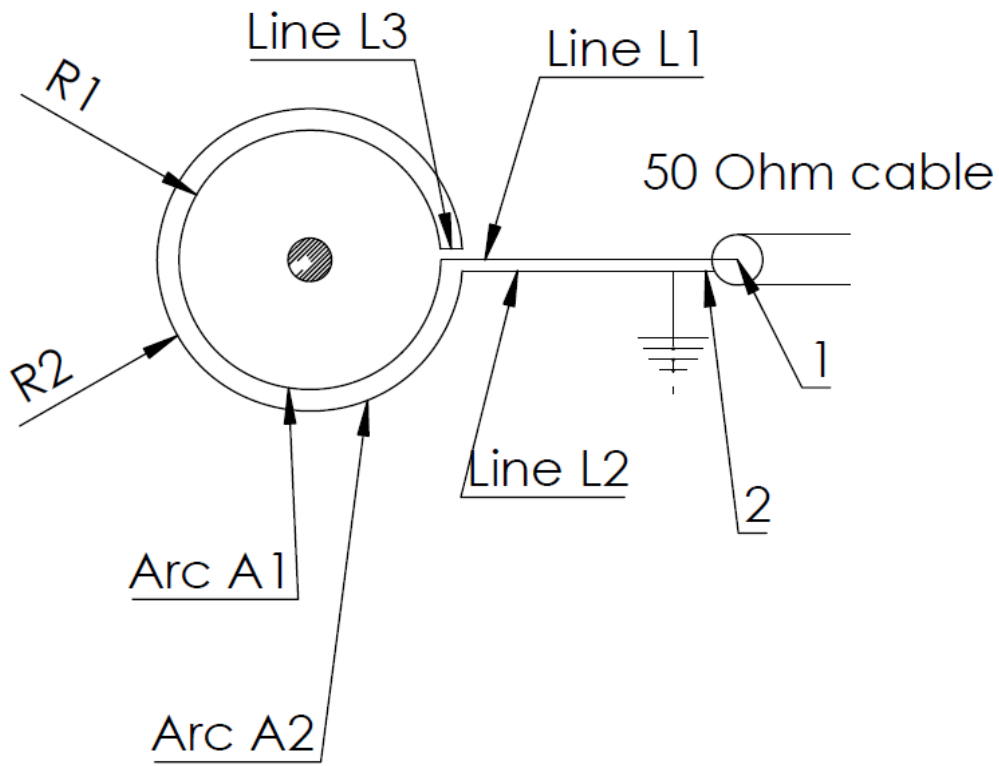

Fig 2: The double loop in relation to the coaxial capacitive voltage divider.

The configurations of Fig 1 and Fig 2 are identical in respect to the displacement current that is induced to flow in the cable, in the limit that the cut portion is a small fraction of the circumference of the coaxial capacitive probe. When the high voltage conductor is a source of only a radial electric field, say when it is a metal wire that is connected to a high voltage capacitor discharge circuit, the two configurations in Fig 1 and Fig 2 should produce exactly the same signal.

But now, consider that the high voltage conductor is a helical wire, like a spring, connected to the high voltage capacitor discharge circuit. Now there is 3-D distribution of current density. In such case, the textbook method of calculating the electric field is to first determine the magnetic vector potential $\vec{A}$ using the formula (23) again for non-radiating fields:

$$\vec{A}(\vec{R},t) = \frac{\mu_0}{4\pi} \int d^3\vec{R}' \frac{\vec{J}(\vec{R}',t')}{|\vec{R}-\vec{R}'|} \tag{23}$$

and then determine the electric field from the relation

$$\vec{E} = -\vec{\nabla}\phi - \frac{\partial \vec{A}}{\partial t} \tag{24}$$

The line integral of the electric field along the contour shown schematically in Fig 2 can be expressed as:

$$V_{1,2} = \int_C \vec{E}.d\vec{\ell} = \int_{L1} \vec{E}.d\vec{\ell} + \int_{A1} \vec{E}.d\vec{\ell} + \int_{L3} \vec{E}.d\vec{\ell} + \int_{A2} \vec{E}.d\vec{\ell} + \int_{L2} \vec{E}.d\vec{\ell} \tag{25}$$

If the gap between the linear segments is chosen to be sufficiently small, the sum of line integrals over the linear segments cancels out. The line integral of electric field along the contour is then equal to

$$V_{1,2} = -\partial_t \left( \int_0^{2\pi-\delta} A_\theta(R_2,\theta,z,t) R_2 d\theta - \int_0^{2\pi-\delta} A_\theta(R_1,\theta,z,t) R_1 d\theta \right) \tag{26}$$

The voltage on the cable is measured assuming the inner conductor to be at positive polarity with respect to the outer conductor as a matter of convention. The inner arc in Fig 2, that is connected to the inner conductor of the cable, traverses the axis in the clockwise ($-\hat{\theta}$) direction while going from the inner cable connection to the outer. Hence the integral over the azimuthal angle has a negative sign for the inner contour.

Here $\delta \ll 2\pi$ is the angular gap created by the cut. In this equation, the contour is idealized as a curve lying in a plane perpendicular to the axis at axial coordinate z. By the circulation theorem, $V_{1,2}$ is the rate of change of poloidal magnetic flux passing through the area $\approx 2\pi R_1 (R_2 - R_1)$ enclosed by this contour as can be verified directly by applying (2). The cylinders forming the coaxial capacitive voltage probe can be idealized as a stack of such contours translated along the axial coordinate. The effective voltage contribution of from the poloidal magnetic flux (PMF) is then an average of $V_{1,2}$ over both the axial and azimuthal coordinates. Following the discussion leading to (22), this contribution is

$$V_{PMF} \equiv \frac{1}{w} \int_{-w/2}^{w/2} dz V_{1,2}$$

$$= -\left( \int_0^{2\pi-\delta} R_2 d\theta \frac{1}{w} \int_{-w/2}^{w/2} dz \partial_t A_\theta(R_2,\theta,z,t) - \int_0^{2\pi-\delta} d\theta R_1 \frac{1}{w} \int_{-w/2}^{w/2} dz \partial_t A_\theta(R_1,\theta,z,t) \right) \quad (27)$$

$$= -\frac{\mu_0}{4\pi} \int dr' \int r' d\theta' \int dz' \partial_t J_\theta(r',\theta',z',t) \left( 2\pi R_2 \mathbb{F}(R_2,r',\theta',z') - 2\pi R_1 \mathbb{F}(R_1,r',\theta',z') \right)$$

It is quite clear that azimuthal variations in $\partial_t J_\theta(r',\theta',z',t)$ are not reflected in $V_{PMF}$

The voltage across the cable input is then the sum of voltages induced by the charge on the inner cylinder given by (14) and (22), and the line integral of electric field given by (26), such voltages being the average over all the axially-stacked contours that make up the cylinders:

$$V_{Cable} = C_1 R_Z \frac{dV_0}{dt} + V_{PMF} \quad (28)$$

When two such double loops are placed near each other, one with the inner conductor clockwise (CW) connected and the other connected counter-clockwise (CCW), the two corresponding voltages will be

$$V_{CCW} = C_1 R_Z \frac{dV_0}{dt} + V_{PMF}; V_{CW} = C_1 R_Z \frac{dV_0}{dt} - V_{PMF} \quad (29)$$

Making symmetric and anti-symmetric combinations of these voltages, one can separate the two contributions

$$C_1 R_Z \frac{dV_0}{dt} = \frac{1}{2}(V_{CCW} + V_{CW}) \equiv V_{SYM}$$
$$V_{PMF} = \frac{1}{2}(V_{CCW} - V_{CW}) \equiv V_{ASYM} \quad (30)$$

While $V_{ASYM}$ is dependent upon the rate of change of azimuthal current density, $V_{SYM}$ is dependent on the rate of change of charge density. The magnitude of the voltage $V_{ASYM}$ is exactly equal to the rate of change of poloidal magnetic flux passing through the annular gap in the double loop.

If there is no emission of poloidal magnetic flux, the signal $V_{ASYM}$ should be zero within experimental errors (such as electromagnetic noise). Conversely, if the signal $V_{ASYM}$ can be shown to have a correlation with plasma dynamics, it would be a conclusive proof of existence of a time-dependent azimuthal current density which is symmetric about the device axis, which would imply existence of axial magnetic field. This leads to the possibility of investigating the experimental conditions which affect it.

The integral equation (27) relating experimentally measured $V_{PMF}$ to the unknown rate of change of azimuthal current density could in principle be solved by having an appropriate number of simultaneous measurement channels. This could be a topic for a very ambitious doctoral research project.

III. **Practical implementation**:

Practical implementation of the above scheme needs to take into account the fact that in case of a large device, the double-loop represents a shorted microstrip transmission line [10] element of significant length for the nanosecond rise-time measurement scheme. It would therefore be prudent to design it with an impedance which matches the (50±3) Ω coaxial cable being used for measurement. The exact formula [10] for the microstrip line can be approximated by

$$Z = \frac{Z_0}{\sqrt{\varepsilon_{eff}}} \frac{h}{w}, Z_0 = 377 \, \Omega \tag{31}$$

wherein $h = R_2 - R_1$ is the annular gap between in the double loop and w is the width of the microstrip. The expression [10] for the effective dielectric constant $\varepsilon_{eff}$ contains the dimensions of the microstrip line also.

A step-by-step approach [8] is described below for constructing this diagnostic taking the UNU-ICTP plasma focus, with outer diameter of cathode about 80 mm as an example. For a first experiment on such a small plasma focus, one could skip the impedance matching steps as that would introduce errors outside the measurement bandwidth, although it would be a good practice not to skip them.

1. Take a plastic tube that would go over the cathode as shown below. Find out how to centre it with respect to the axis. Perhaps put suitable packing between the tube and the cathode. One could also use a roll of some stiff plastic sheet such as overhead transparency sheet. The ultimate objective is to support two double loops over that tube, at the elevation of the pinch as shown in Fig 2.

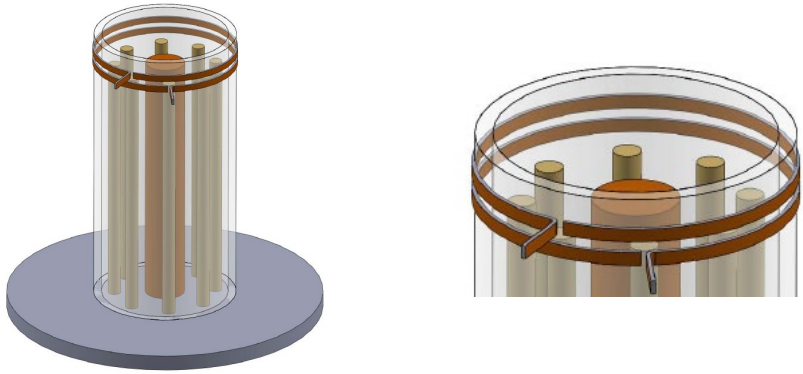

Fig 3: Plastic tube that goes over the cathode and serves as a support for the CCW and CW double loops.

2. Find out whether you can buy copper tape such as (https://www.amazon.com/Sparkfun-Electronics-PRT-10561-Copper-Tape/dp/B007R9UOBM). Width could be from 5-20 mm. It will be good if it has adhesive on one side with a protective liner over it.

3. Use double-sided foam adhesive tape as dielectric. Dielectric constant of this tape is aprox. 2. Data on 3M VHB Foam Tapes is reproduced below:

| | 3M™ VHB™ Tape 4941 | 3M™ VHB™ Tape 5952 | Units | Test Standard |
|---|---|---|---|---|
| Dielectric Constant | 2.29 / 1.99 | 2.14 / 1.95 | at 1 kHz / at 1MHz | ASTM D150 / ASTM D150 |
| Dissipation Factor | 0.0245 / 0.0374 | 0.0065 / 0.0506 | at 1 kHz / at 1MHz | ASTM D150 / ASTM D150 |
| Dielectric Breakdown Strength | 14 (360) | 18 (455) | V/μm (V/mil) | ASTM D140 |
| Thermal Conductivity (k value) | 0.08 (0.5) | 0.05 (0.4) | W/mK (BTU•in/hr•ft²•°F) | |
| Volume Resisitivity | 2.1 x 10¹⁴ | 2.5 x 10¹⁴ | Ω-cm | ASTM D257 |
| Surface Resisitivity | 2.7 x 10¹⁴ | >10¹⁶ | Ω/sq | ASTM D257 |
| Water Vapor Transmission Rate | 25.6 | 37.1 | g/m²•day | ASTM F1249 at 38°C/100% RH |

| Tape Number | Color | Thickness in (mm) |
|---|---|---|
| 4919F | Black | 0.025 (0.6) |
| 4926 | Gray | 0.015 (0.4) |
| 4936(F) | Gray | 0.025 (0.6) |
| 4941(F) | Gray | 0.045 (1.1) |
| 4947F | Black | 0.045 (1.1) |
| 4956(F) | Gray | 0.062 (1.6) |
| 4979F | Black | 0.062 (1.6) |
| 4991 | Gray | 0.090 (2.3) |
| 4991B | Black | 0.090 (2.3) |

| Tape Number | Color | Thickness in (mm) |
|---|---|---|
| 5906 | Black | 0.006 (0.15) |
| 5907 | Black | 0.008 (0.20) |
| 5908 | Black | 0.010 (0.25) |
| 5909 | Black | 0.012 (0.30) |
| 5915(P) | Black | 0.016 (0.4) |
| 5915WF | White | 0.016 (0.4) |
| 5925(P) | Black | 0.025 (0.6) |
| 5925WF | White | 0.025 (0.6) |
| 5930(P) | Black | 0.032 (0.8) |
| 5930WF | White | 0.032 (0.8) |
| 5952(P) | Black | 0.045 (1.1) |
| 5952WF | White | 0.045 (1.1) |
| 5958FR | Black | 0.040 (1.0) |
| 5962(P) | Black | 0.062 (1.6) |
| 5962WF | White | 0.062 (1.6) |

4. The formula given in the reference [10] leads to the following values of w/h for ε=1.99

| Z   | 44  | 45   | 46   | 47   | 48   | 49   | 50   | 51   | 52   | 53   | 54   | 55   |
|-----|-----|------|------|------|------|------|------|------|------|------|------|------|
| w/h | 3.9 | 3.78 | 3.66 | 3.55 | 3.44 | 3.34 | 3.24 | 3.15 | 3.06 | 2.97 | 2.89 | 2.81 |

5. If you have a capacitance meter, you can measure the impedance of the coaxial cable you are using the following steps. Measure the length X of the cable in meters. Measure its capacitance C in nano-farads. Then the impedance is 5X/C in Ohms. The above table then gives the value of w/h needed to match the microstrip line impedance with that of the cable. (This step may be skipped for an initial experiment. It a desirable but not-absolutely-necessary step.)

6. Measure (or find from manufacturer's data) the thickness h of the double-sided tape. Multiple layers could also be used in case the thickness is too small.

7. Measure the circumference S of the plastic tube in mm. Cut the copper tape to a length of 2*S + 40 mm. Cut the double-sided foam adhesive tape to a length of S+ 20 mm. Fold the copper tape lengthwise over the double-sided tape, with the liner of double-sided tape removed, with the adhesive side of copper tape on the outside. The fold corresponds to Line L3 in Fig 2.

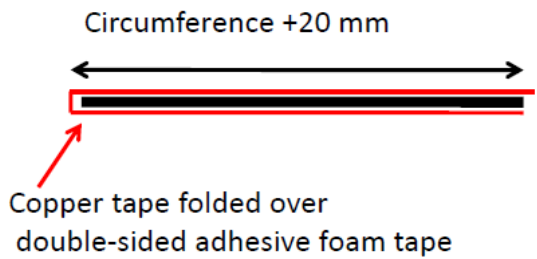

Fig 4: Illustration of above text

8. Cut this sandwich length-wise using a sharp knife and ruler to a width w calculated using the thickness of the double-sided tape in step 6 and value of w/h determined in step 4. This becomes one of the two coils. If the sandwich is wide enough, cut another strip to make the second coil. At this stage the adhesive on the copper tape is still covered with its protective liner, which is facing outwards in Fig 4.

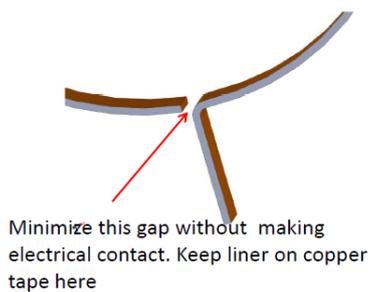

Fig 5: Illustration of above text

9. Mark the position of top of anode on the plastic tube. Peel the liner on the copper tape from one side and cut it carefully making sure that it remains over the fold. Wrap it on the plastic tube above the mark and press to stick it. When it goes completely around and reaches the fold corresponding to Line L3 in Fig 2, bend the sandwich 90 degrees to form a configuration similar to Fig 2. Secure the two ends of the wrapped transmission line to the plastic tube using scotch tape. The portion of liner on copper tape covering the fold corresponding to Line L3 should determine the angular gap δ.

10. Note the sense (Clockwise CW or Counter Clockwise CCW) of the first transmission line. Wrap the second one in a similar manner with a gap of 2 mm with the opposite sense. Note that in the cylindrical coordinate system with the z-axis coincident with the device axis, the CCW orientation is along +θ.

11. Solder a short length of coaxial cable to the 20 mm stub to take the signal out. The side of the copper strip facing the plasma must be soldered to the inner cable conductor. Inspect the solder joint using a magnifier to ensure that there is no accidental short circuit across the cable via any loose copper strand or a solder bead. After inspection, consider securing and insulating the solder joint using a glue gun or a quick-setting adhesive such as cyanoacrylate. The other end of the coaxial cable should be connectorized appropriately prior to this step to enable the signal to be taken out of the vacuum chamber.

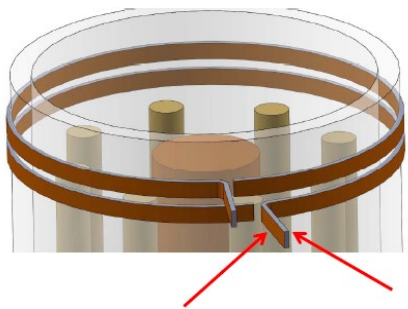

Fig 6: Illustration of the text.

Coaxial cable gets soldered on two sides of the sandwich micro-strip line

## IV. <u>Testing, validation and calibration:</u>

Several reasonable doubts must be addressed using proper experimental procedures, which are discussed below.

A. The adhesive strip may have a frequency-dependent dielectric constant. Does it have any effect on the frequency response of the diagnostic that may lead to mis-interpretation?

The best experimental test of this question is to measure the square-pulse response of the diagnostic. This is achieved by setting up the following arrangement using standard BNC cables and connectors:

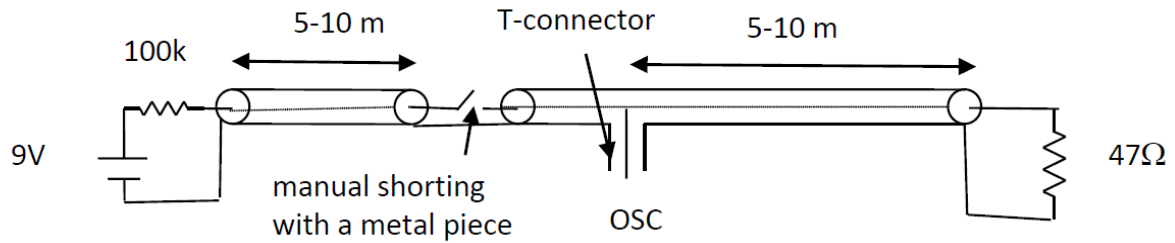

a) The first cable is charged to 9V using a battery. On manual shorting, this cable discharges into the second cable. The discharge is completed in a time $t_1$=10 ns * Length of 1st cable in meters. The OSC sees a square pulse of width $t_1$ and height 4.5 V. When the second cable is terminated in 47Ω, there is no reflection and only one pulse is seen.

b) Remove the 47 Ω and do this again. Now the reflection coefficient is 1 and the first pulse is followed by several reflected pulses. The separation between the first pulse and the first reflected pulse is $t_2$= 10ns * Length of second cable in meters. The change in shape between the first pulse and the first reflected pulse, if any, is because of bandwidth of the cable. If the cable is of inferior quality, for example using recycled dielectric, there may be severe distortion.

c) Now replace the 47 Ω with a short circuit. The reflection coefficient is now -1 and the polarity of the pulse is reversed. But the shape should be a mirror image of the open circuit reflected pulse.

d) Now replace the short circuit with the loop made with copper tape and double-sided adhesive tape. The difference in shape between the reflected pulse of this test and the short circuit test is because of the dielectric frequency response of the double-sided adhesive tape. If there is negligible difference, it means that the frequency response of the dielectric does not have any noticeable effect on the measurement.

B. Is the $V_{ASYM}$ signal non-zero even if there is no azimuthal current density?

Theoretical considerations do suggest that the $V_{ASYM}$ signal should be zero when there is no azimuthal current density. But there could be a reasonable doubt whether the experimental arrangement actually reproduces the assumed configuration. A simple test can be organized as follows. The following household electric components are wired in series: a plug, a fan capacitor (or a tungsten filament lamp) and an electronic fan regulator with sufficient free length of insulated wire. A tight U-bend is made on the insulated wire and secured with adhesive tape to a plastic rod. Support the plastic rod over the anode. Now put the plug in a socket and switch on. The electronic fan regulator repeatedly breaks the circuit causing voltage and current transients. Since the current distribution has only axial and radial components, the $V_{ASYM}$ signal should be zero. In practice, it may have a small, nonzero value because of inadequate shielding of the solder joint. Some kind of metallic shielding which does not short the cable can be tried using copper tape and glue gun. The $V_{SYM}$ signal, on the other hand, should prominently display the transients introduced by the electronic fan regulator.

By placing the rod in off-axis and inclined positions, the sensitivity of the $V_{ASYM}$ signal to geometrical deviations from symmetry can be mapped in principle.

C. Would a zero signal in $V_{ASYM}$ reliably imply that there is no azimuthal current density in the plasma?

A zero signal in $V_{ASYM}$ could also arise through a short-circuit across the cable. To ensure that that is not the case, make a helix of the insulated wire around the plastic rod and test again. There should be similar-shaped signals in both $V_{ASYM}$ and $V_{SYM}$ showing transients. This would rule out a short circuit across the cable. If this test is positive and there is no $V_{ASYM}$ signal in a plasma shot, it reliably indicates absence of azimuthal current density in the plasma. If a $V_{ASYM}$ signal is observed in a plasma shot, it should have some structure that correlates with known plasma phenomena such as the current derivative singularity, neutron and x-ray signals etc. Such correlated structure would conclusively establish presence of an axis-symmetric component of azimuthal current density in the plasma.

D. Is it possible to have any kind of calibration?

A calibrated measurement of any plasma-related quantity is not within the intended scope of this diagnostic. Nevertheless, some ingenious efforts can be made towards some calibration, with no guarantee of success. It could involve making a conducting mock-up of the plasma in the pinch phase using 3-D printing technology, which is placed on the anode. The voltage transient applied to this mock-up using the electronic fan regulator could be simultaneously measured using a calibrated differential voltage probe. This would provide some calibration of the $V_{SYM}$ signal in terms of the dV/dt at the anode tip according to (30). The numerical integration of the $V_{SYM}$ signal could then be related to the pinch voltage, assuming that the voltage offset in the digital oscilloscope channel is either known or zero.

The $V_{ASYM}$ signal is numerically equal to the rate of change of poloidal magnetic flux passing through the annulus. Measurement of the annular gap provides a calibration of the average poloidal magnetic field in that gap. This could be compared with models of azimuthal current distribution in the plasma.

V. Summary and conclusion:

This paper introduces a new diagnostic technique that can be applied to a small plasma focus machine with modest resources. It provides an unambiguous diagnostic about the existence of poloidal magnetic flux emission from the plasma focus: an area not much researched so far. It produces two signals: one proportional to the rate of change of charge on the plasma focus and the other, proportional to the rate of change of azimuthal current density.

It has been tested on a plasma focus in Sofia, Bulgaria and the results are being presented elsewhere. It reveals new information about the plasma focus not obtainable by other techniques.

It is hoped that this technique will be fielded, explored and improved and studied in correlation with other plasma focus diagnostics such as x-rays and neutrons by young plasma focus researchers, who are welcome to contact the author for help and support.